\documentstyle[12pt]{article}

\catcode`\@=11
\long\def\@makefntext#1{ 
\protect\noindent \hbox to 3.2pt {\hskip-.9pt
$^{{\ninerm\@thefnmark}}$\hfil}#1\hfill} 

\def\thefootnote{\fnsymbol{footnote}}
 \def\@makefnmark{\hbox to 0pt{$^{\@thefnmark}$\hss}}  

\def\ps@myheadings{\let\@mkboth\@gobbletwo
\def\@oddhead{\hbox{} 
\rightmark\hfil\ninerm\thepage}
\def\@oddfoot{}\def\@evenhead{\ninerm\thepage\hfil 
\leftmark\hbox{}}\def\@evenfoot{}
\def\sectionmark##1{}\def\subsectionmark##1{}}

\begin{document}

\newcommand{\symbolfootnote}{\renewcommand{\thefootnote}
        {\fnsymbol{footnote}}}
\renewcommand{\thefootnote}{\fnsymbol{footnote}}
\newcommand{\alphfootnote}
        {\setcounter{footnote}{0}
         \renewcommand{\thefootnote}{\sevenrm\alph{footnote}}}

\newcounter{sectionc}\newcounter{subsectionc}\newcounter{subsubsectionc}
\renewcommand{\section}[1] {\vspace{0.6cm}\addtocounter{sectionc}{1}
\setcounter{subsectionc}{0}\setcounter{subsubsectionc}{0}\noindent
        {\bf\thesectionc. #1}\par\vspace{0.4cm}}
\renewcommand{\subsection}[1] {\vspace{0.6cm}\addtocounter{subsectionc}{1}
        \setcounter{subsubsectionc}{0}\noindent
        {\it\thesectionc.\thesubsectionc. #1}\par\vspace{0.4cm}}
\renewcommand{\subsubsection}[1]
{\vspace{0.6cm}\addtocounter{subsubsectionc}{1}
        \noindent {\rm\thesectionc.\thesubsectionc.\thesubsubsectionc.
        #1}\par\vspace{0.4cm}}
\newcommand{\nonumsection}[1] {\vspace{0.6cm}\noindent{\bf #1}
        \par\vspace{0.4cm}}

\newcounter{appendixc}
\newcounter{subappendixc}[appendixc]
\newcounter{subsubappendixc}[subappendixc]
\renewcommand{\thesubappendixc}{\Alph{appendixc}.\arabic{subappendixc}}
\renewcommand{\thesubsubappendixc}
        {\Alph{appendixc}.\arabic{subappendixc}.\arabic{subsubappendixc}}

\renewcommand{\appendix}[1] {\vspace{0.6cm}
        \refstepcounter{appendixc}
        \setcounter{figure}{0}
        \setcounter{table}{0}
        \setcounter{equation}{0}
        \renewcommand{\thefigure}{\Alph{appendixc}.\arabic{figure}}
        \renewcommand{\thetable}{\Alph{appendixc}.\arabic{table}}
        \renewcommand{\theappendixc}{\Alph{appendixc}}
        \renewcommand{\theequation}{\Alph{appendixc}.\arabic{equation}}
        \noindent{\bf Appendix \theappendixc #1}\par\vspace{0.4cm}}
\newcommand{\subappendix}[1] {\vspace{0.6cm}
        \refstepcounter{subappendixc}
        \noindent{\bf Appendix \thesubappendixc. #1}\par\vspace{0.4cm}}
\newcommand{\subsubappendix}[1] {\vspace{0.6cm}
        \refstepcounter{subsubappendixc}
        \noindent{\it Appendix \thesubsubappendixc. #1}
        \par\vspace{0.4cm}}

\def\abstracts#1{{
        \centering{\begin{minipage}{30pc}\tenrm\baselineskip=12pt\noindent
        \centerline{\tenrm ABSTRACT}\vspace{0.3cm}
        \parindent=0pt #1
        \end{minipage} }\par}}

\newcommand{\bibit}{\it}
\newcommand{\bibbf}{\bf}
\renewenvironment{thebibliography}[1]
        {\begin{list}{\arabic{enumi}.}
        {\usecounter{enumi}\setlength{\parsep}{0pt}
\setlength{\leftmargin 1.25cm}{\rightmargin 0pt}
         \setlength{\itemsep}{0pt} \settowidth
        {\labelwidth}{#1.}\sloppy}}{\end{list}}

\topsep=0in\parsep=0in\itemsep=0in
\parindent=1.5pc

\newcounter{itemlistc}
\newcounter{romanlistc}
\newcounter{alphlistc}
\newcounter{arabiclistc}
\newenvironment{itemlist}
        {\setcounter{itemlistc}{0}
         \begin{list}{$\bullet$}
        {\usecounter{itemlistc}
         \setlength{\parsep}{0pt}
         \setlength{\itemsep}{0pt}}}{\end{list}}

\newenvironment{romanlist}
        {\setcounter{romanlistc}{0}
         \begin{list}{$($\roman{romanlistc}$)$}
        {\usecounter{romanlistc}
         \setlength{\parsep}{0pt}
         \setlength{\itemsep}{0pt}}}{\end{list}}

\newenvironment{alphlist}
        {\setcounter{alphlistc}{0}
         \begin{list}{$($\alph{alphlistc}$)$}
        {\usecounter{alphlistc}
         \setlength{\parsep}{0pt}
         \setlength{\itemsep}{0pt}}}{\end{list}}

\newenvironment{arabiclist}
        {\setcounter{arabiclistc}{0}
         \begin{list}{\arabic{arabiclistc}}
        {\usecounter{arabiclistc}
         \setlength{\parsep}{0pt}
         \setlength{\itemsep}{0pt}}}{\end{list}}

\newcommand{\fcaption}[1]{
        \refstepcounter{figure}
        \setbox\@tempboxa = \hbox{\tenrm Fig.~\thefigure. #1}
        \ifdim \wd\@tempboxa > 6in
           {\begin{center}
        \parbox{6in}{\tenrm\baselineskip=12pt Fig.~\thefigure. #1 }
            \end{center}}
        \else
             {\begin{center}
             {\tenrm Fig.~\thefigure. #1}
              \end{center}}
        \fi}

\newcommand{\tcaption}[1]{
        \refstepcounter{table}
        \setbox\@tempboxa = \hbox{\tenrm Table~\thetable. #1}
        \ifdim \wd\@tempboxa > 6in
           {\begin{center}
        \parbox{6in}{\tenrm\baselineskip=12pt Table~\thetable. #1 }
            \end{center}}
        \else
             {\begin{center}
             {\tenrm Table~\thetable. #1}
              \end{center}}
        \fi}

\def\@citex[#1]#2{\if@filesw\immediate\write\@auxout
        {\string\citation{#2}}\fi
\def\@citea{}\@cite{\@for\@citeb:=#2\do
        {\@citea\def\@citea{,}\@ifundefined
        {b@\@citeb}{{\bf ?}\@warning
        {Citation `\@citeb' on page \thepage \space undefined}}
        {\csname b@\@citeb\endcsname}}}{#1}}

\newif\if@cghi
\def\cite{\@cghitrue\@ifnextchar [{\@tempswatrue
        \@citex}{\@tempswafalse\@citex[]}}
\def\citelow{\@cghifalse\@ifnextchar [{\@tempswatrue
        \@citex}{\@tempswafalse\@citex[]}}
\def\@cite#1#2{{$\null^{#1}$\if@tempswa\typeout
        {IJCGA warning: optional citation argument
        ignored: `#2'} \fi}}
\newcommand{\citeup}{\cite}

\def\fnm#1{$^{\mbox{\scriptsize #1}}$}
\def\fnt#1#2{\footnotetext{\kern-.3em
        {$^{\mbox{\sevenrm #1}}$}{#2}}}

\font\twelvebf=cmbx10 scaled\magstep 1
\font\twelverm=cmr10 scaled\magstep 1
\font\twelveit=cmti10 scaled\magstep 1
\font\elevenbfit=cmbxti10 scaled\magstephalf
\font\elevenbf=cmbx10 scaled\magstephalf
\font\elevenrm=cmr10 scaled\magstephalf
\font\elevenit=cmti10 scaled\magstephalf
\font\bfit=cmbxti10
\font\tenbf=cmbx10
\font\tenrm=cmr10
\font\tenit=cmti10
\font\ninebf=cmbx9
\font\ninerm=cmr9
\font\nineit=cmti9
\font\eightbf=cmbx8
\font\eightrm=cmr8
\font\eightit=cmti8


\centerline{\tenbf QUANTUM ALGEBRAIC SYMMETRIES
  IN NUCLEAR AND MOLECULAR PHYSICS}
\vspace{0.8cm}
\centerline{\tenrm Dennis BONATSOS}
\baselineskip=13pt
\centerline{\tenit ECT$^*$, Villa Tambosi, Strada delle Tabarelle 286}
\baselineskip=12pt
\centerline{\tenit I-38050 Villazzano (Trento), Italy}
\vspace{0.3cm}
\centerline{\tenrm C. DASKALOYANNIS}
\baselineskip=13pt
\centerline{\tenit Department of Physics, Aristotle University of Thessaloniki}
\baselineskip=12pt
\centerline{\tenit GR-54006 Thessaloniki, Greece}
\vspace{0.3cm}
\centerline{\tenrm P. KOLOKOTRONIS, D. LENIS}
\baselineskip=13pt
\centerline{\tenit Institute of Nuclear Physics, NCSR ``Demokritos''}
\baselineskip=12pt
\centerline{\tenit GR-15310 Aghia Paraskevi, Attiki, Greece}
\vspace{0.9cm}
\abstracts{Various applications of quantum algebraic techniques in nuclear
structure physics and in molecular physics are briefly reviewed.}

\vfil
\rm\baselineskip=14pt
\section{Introduction}

Quantum algebras (also called quantum groups) are deformed versions
of the
usual Lie algebras, to which they reduce when the deformation parameter
$q$ is set equal to unity. From the mathematical point of view they are
Hopf algebras.
Their use in physics became popular with the
introduction $^{1,2}$ of the $q$-deformed harmonic oscillator as a tool for
providing a boson realization of the quantum algebra su$_q$(2), although
similar mathematical structures had already been known $^{3}$.
Initially used for solving the quantum Yang--Baxter equation,
quantum algebras
have subsequently found applications in several branches of physics, as, for
example, in the description of spin chains, squeezed states
$^{4}$, hydrogen atom and hydrogen-like spectra $^{5-7}$
rotational
and vibrational nuclear and molecular spectra
and in conformal
field theories. By now much work has been done $^{8-11}$
on the $q$-deformed oscillator and its relativistic extensions $^{12,13}$, and
several kinds of generalized deformed oscillators
$^{14-16}$
and generalized deformed su(2) algebras $^{17,18}$  have been introduced.

Here we shall confine ourselves to applications of quantum algebras in nuclear
structure physics and in molecular physics. The purpose of this
short review is to provide the  reader with references for further reading.

\section{ The su$_q$(2) rotator model}

 The first application of quantum algebras in nuclear physics was the use
of the deformed algebra su$_q$(2) for the description of the rotational
spectra of deformed $^{19,20}$ and superdeformed $^{21}$ nuclei.
The Hamiltonian of the $q$-deformed rotator is proportional to the
second order Casimir operator of the su$_q$(2) algebra. Its Taylor expansion
contains powers of $J(J+1)$ (where $J$ is the angular momentum), being
similar $^{20}$ to the expansion provided by the Variable Moment of Inertia
(VMI) model. Furthermore, the deformation
parameter $\tau$ (with $q=e^{i\tau}$) has been found $^{20}$ to correspond to
the softness parameter of the VMI model. Through a comparison
of the su$_q$(2) model to the hybrid model
the deformation parameter $\tau$
has also been connected to the number of valence nucleon pairs $^{22}$
and to the nuclear deformation $\beta$ $^{23}$. Since $\tau$ is an indicator
of deviation from the pure su(2) symmetry, it is not surprising that
$\tau$ decreases with increasing $\beta$ $^{23}$.

B(E2) transition probabilities have also been described in this framework
$^{24}$.
In this case the $q$-deformed Clebsch--Gordan coefficients are used instead
of the normal ones. (It should be noticed that the $q$-deformed angular
momentum theory has already been much developed $^{24}$.) The model predicts
an increase of the B(E2) values with angular momentum, while the rigid
rotator model predicts saturation. Some experimental results supporting
this prediction already exist $^{24}$. Similarly increasing B(E2) values
are predicted by a modified version $^{25}$ of the su(3) limit of the
Interacting Boson Model (IBM), by the Fermion Dynamical Symmetry
Model (FDSM)
$^{26}$, as well as by the recent
systematics of Zamfir and Casten $^{27}$.

\section{ Extensions of the su$_q$(2) model}

The su$_q$(2) model has been successful in describing rotational nuclear
spectra. For the description of vibrational and transitional nuclear
spectra it has been found $^{28}$ that $J(J+1)$ has to be replaced by $J(J+c)$.
The additional parameter $c$ allows for the description of nuclear
anharmonicities in a way similar to that of the Interacting Boson Model
(IBM) and the Generalized Variable Moment of Inertia (GVMI) model
$^{29}$. The use of $J(J+c)$ instead of $J(J+1)$ for vibrational and
transitional nuclei is also supported by recent systematics $^{30}$.

Another generalization is based on the use of the deformed algebra
su$_{\Phi}$(2) $^{17,18}$, which is characterized by a structure function
$\Phi$.
The usual su(2) and su$_q$(2) algebras are obtained for specific choices
of the structure function $\Phi$. The su$_{\Phi}$(2) algebra has been
constructed so that its representation theory resembles as much as possible
the representation theory of the usual su(2) algebra. Using this technique
one can construct, for example, a rotator having the same spectrum as the
one given by the Holmberg--Lipas formula $^{31}$.
A two-parameter generalization of the su$_q$(2) model, labelled as
su$_{qp}$(2), has also been successfully used for the description of
superdeformed nuclear bands $^{32}$.

\section{ Pairing correlations}

It has been found $^{33}$ that correlated fermion pairs coupled to zero
angular
momentum in a single-$j$ shell behave approximately as suitably defined
$q$-deformed bosons. After performing the same boson mapping to a simple
pairing Hamiltonian, one sees that the pairing energies are also correctly
reproduced up to the same order. The deformation parameter used ($\tau
=\ln q$) is found to be inversely proportional to the size of the shell,
thus serving as a small parameter.

The above mentioned system of correlated fermion pairs can be described
{\sl exactly} by suitably defined generalized deformed bosons $^{34}$. Then
both the commutation relations are satisfied exactly and the pairing energies
are reproduced exactly. The spectrum of the appropriate generalized
deformed oscillator corresponds, up to first order perturbation theory,
to a harmonic oscillator with an $x^4$ perturbation.

If one considers, in addition to the pairs coupled to zero angular momentum,
pairs coupled to non-zero angular momenta, one finds that an approximate
description in terms of two suitably defined $q$-oscillators (one describing
the $J=0$ pairs and the other corresponding to the $J\neq 0$ pairs) occurs
$^{35}$. The additional terms introduced by the deformation have been found
$^{35}$ to improve the description of the neutron pair separation energies
of the Sn isotopes, with no extra parameter introduced.

$q$-deformed versions of the pairing theory have also been given in
$^{36,37}$.

\section{ $q$-deformed versions of nuclear models}

A $q$-deformed version of a two dimensional toy Interacting Boson Model (IBM)
 with su$_q$(3) overall symmetry
has been developed $^{38,39}$, mainly for testing the ways in which
spectra and transition probabilities are influenced by the $q$-deformation.
The question of possible complete breaking of the symmetry through
$q$-deformation, i.e. the transition from the su$_q$(2) limiting symmetry
to the so$_q$(3) one has been examined $^{40, 41}$. It has been found that
such a transition is possible for complex values of the parameter $q$ $^{41}$.
(For problems arising when using complex $q$ values see $^{42}$).
Complete breaking of the symmetry has also been considered in the framework
of an su$_q$(2) model $^{43}$. It has also been found $^{44}$ that
$q$-deformation leads (for specific range of values of the deformation
parameter $\tau$, with $q=e^{i\tau}$) to a recovery of the u(3) symmetry
in the framework of a simple Nilsson model including a spin-orbit term.
Finally, the o$_q$(3) limit of the toy IBM model has been used for the
description of $^{16}$O + $\alpha$ cluster states in $^{20}$Ne, with
positive results $^{45}$.

$q$-deformed versions of the o(6) and u(5) limits of the full IBM have been
discussed in $^{46-48}$. The $q$-deformation of the su(3) limit of IBM
is a formidable problem, since the su$_q$(3) $\supset$ so$_q$(3)
decomposition has for the moment been achieved only for completely symmetric
su$_q$(3) irreducible representations $^{49}$.

Furthermore a $q$-deformed version of the Moszkowski model
has been developed
$^{50}$ and RPA modes have been studied $^{51}$ in it. A $q$-deformed
Moszkowski model
with cranking has also been studied $^{52}$ in the mean-field approximation.
It has been seen that the residual interaction simulated by the
$q$-deformation is felt more strongly by states with large $J_z$. The
possibility of using $q$-deformation in assimilating temperature effects is
receiving attention, since it has also been found $^{53}$ that this approach
can be used in describing thermal effects in the framework of a $q$-deformed
Thouless model for supercoductivity.

In addition, $q$-deformed versions of the Lipkin-Meshkov-Glick (LMG)
model have been developed, both for the 2-level version of the model
in terms of an su$_q$(2) algebra $^{54}$, and for the 3-level version
of the model in terms of an su$_q$(3) algebra $^{55}$.

 \vfill\eject

\section{ Anisotropic quantum harmonic oscillator with rational ratios of
frequencies}

The symmetries of the 3-dimensional anisotropic quantum harmonic oscillator
with rational ratios of frequencies (RHO) are of high current interest in
nuclear physics, since they are
the basic symmetries underlying the structure of superdeformed and
hyperdeformed nuclei.
The 2-dimensional RHO is also of interest, in
connection with ``pancake'' nuclei, i.e. very oblate nuclei. Cluster
configurations in light nuclei can also be  described in terms of RHO
symmetries, which underlie the geometrical structure of the
Bloch--Brink $\alpha$-cluster model. The 3-dim RHO is also of
interest for the interpretation
of the observed shell structure in atomic clusters, especially
after the
realization that large deformations can occur in such systems.

The two-dimensional  and
three-dimensional $^{56}$  anisotropic harmonic
oscillators have been the subject of several investigations, both at the
classical and the quantum mechanical level (see $^{57,58}$ for references).
These oscillators are examples
of superintegrable systems. The special cases with frequency
ratios 1:2 and 1:3 have also been considered $^{59}$. While
at the classical level it is clear that the su(N) or sp(2N,R) algebras can
be used for the description of the N-dimensional anisotropic oscillator, the
situation at the quantum level, even in the two-dimensional case, is not as
simple.
It has been proved that a generalized deformed u(2)
algebra is the symmetry algebra of the two-dimensional anisotropic quantum
harmonic oscillator $^{57}$, which is the oscillator describing the
single-particle
level spectrum of ``pancake'' nuclei, i.e. of triaxially deformed nuclei
with $\omega_x >> \omega_y$, $\omega_z$. Furthermore, a generalized
deformed u(3) algebra turns out to be the symmetry algebra of the
three-dimensional RHO $^{58}$.

\section{The use of quantum algebras in molecular structure}

Similar techniques can be applied
in describing properties of diatomic and polytomic molecules. A brief
list will be given here.

1) Rotational spectra of diatomic molecules have been described in terms of
the su$_q$(2) model $^{60}$.
 As in the case of nuclei, $q$ is a phase factor
($q=e^{i\tau}$). In molecules $\tau$ is of the order of 0.01.
The use of the su$_q$(2) symmetry leads to a partial summation of the Dunham
expansion describing the rotational--vibrational spectra of diatomic
molecules $^{60}$. Molecular backbending (bandcrossing) has also been
described in this framework $^{61}$. Rotational spectra of symmetric
top molecules have also been considered $^{62,63}$ in the
framework of the su$_q$(2) symmetry.

2) Vibrational spectra of diatomic molecules have been described in terms of
$q$-deformed anharmonic oscillators having the
su$_q$(1,1) $^{64}$ or the u$_q$(2) $\supset$ o$_q$(2) $^{65}$
symmetry, as well as in terms of generalized deformed oscillators
similar to the ones described in sec. 3 $^{66,67}$.
These results, combined with 1), lead
to the full summation of the Dunham expansion $^{64,65}$.
A two-parameter deformed anharmonic oscillator with u$_{qp}$(2) $\supset$
o$_{qp}$(2) symmetry has also been considered $^{68}$.

3) The physical content of the anharmonic oscillators mentioned in 2)
has been clarified by constructing WKB equivalent potentials (WKB-EPs)
and classical equivalent potentials $^{69}$ providing approximately the same
spectrum.
The results have been corroborated by the study of the
relation between su$_q$(1,1) and the anharmonic oscillator with  $x^4$
anharminicities $^{70}$. Furthermore
the WKB-EP corresponding to the su$_q$(1,1) anharmonic
oscillator has been connected to a class of Quasi-Exactly Soluble Potentials
(QESPs) $^{71}$.

4) Generalized deformed oscillators
giving the same spectrum as the Morse potential $^{72}$ and the modified
P\"oschl--Teller potential $^{73}$,  as well as a deformed oscillator
containing them as special cases $^{74}$ have also been
constructed.
In addition,  $q$-deformed versions of the Morse potential have been given,
either by using the so$_q$(2,1) symmetry $^{75}$ or by solving a
$q$-deformed Schr\"odinger equation for the usual Morse potential
$^{76}$.

5) A $q$-deformed version of the vibron model for diatomic molecules has been
constructed $^{77}$, in a way similar to that described in sec. 5.

6) For vibrational spectra of polyatomic molecules a model of $n$ coupled
generalized deformed oscillators has been built $^{78}$, containg the
approach of Iachello and Oss $^{79}$ as a special case.

7) Quasi-molecular resonances in the systems $^{12}$C+$^{12}$C and
$^{12}$C+$^{16}$O have been described in terms of a $q$-deformed oscillator
plus a rigid rotator $^{80}$.

A review of several of the above topics, accompanied by a detailed and
self-contained introduction to quantum algebras, has been given by Raychev
$^{81}$.

\end{document}